\documentclass{article} 

\usepackage[utf8]{inputenc} 

\twocolumn

\usepackage{graphicx} 

\usepackage{booktabs} 
\usepackage{array} 
\usepackage{paralist} 
\usepackage{verbatim} 

\usepackage{hyperref}

\begin{document}
\title{\makebox[\textwidth]{Induction of slow oscillations by rhythmic acoustic stimulation}}
\author{Hong-Viet V. Ngo$^{1,2}$, Jens Christian Claussen$^1$,  Jan Born$^{3,4}$ and Matthias  M\"olle$^{3,4}$
\\ \footnotesize $^{1}$Institute for Neuro- and Bioinformatics, University of L\"ubeck, Germany, 
\\ \footnotesize $^2$Graduate School for Computing in Medicine and Life Sciences,
University of L\"ubeck, Germany, 
\\\footnotesize $^3$Department of Neuroendocrinology, University of L\"ubeck, Germany and 4Department of Medical Psychology and 
\\ \footnotesize $^4$Behavioral Neurobiology, University of T\"ubingen, Germany}
\date{6. March 2012}
\maketitle
\noindent
{\footnotesize This is an author-generated file for personal non-commercial\\
 use only. For reference, please refer to the published version at}\\
\fbox{\href{http://dx.doi.org:10.1111/j.1365-2869.2012.01039.x}{J. Sleep Res. (2013), 22, 22-31}}
\\
doi: \href{http://dx.doi.org:10.1111/j.1365-2869.2012.01039.x}{10.1111/j.1365-2869.2012.01039.x}
\setcounter{page}{22}
\begin{abstract}
\large \sf \bf
Slow oscillations are electrical potential oscillations with a spectral peak
frequency of $\sim$0.8 Hz, and hallmark the electroencephalogram during
slow-wave sleep. Recent studies have indicated a causal contribution of
slow oscillations to the consolidation of memories during slow-wave
sleep, raising the question to what extent such oscillations can be
induced by external stimulation. Here, we examined whether slow
oscillations can be effectively induced by rhythmic acoustic stimulation.
Human subjects were examined in three conditions: (i) with tones
presented at a rate of 0.8 Hz (`0.8-Hz stimulation'); (ii) with tones
presented at a random sequence (`random stimulation'); and (iii) with no
tones presented in a control condition  (`sham'). Stimulation started during
wakefulness before sleep and continued for the first $\sim$90 min of sleep.
Compared with the other two conditions, 0.8-Hz stimulation significantly
delayed sleep onset. However, once sleep was established, 0.8-Hz
stimulation significantly increased and entrained endogenous slow
oscillation activity. Sleep after the 90-min period of stimulation did not
differ between the conditions. Our data show that rhythmic acoustic
stimulation can be used to effectively enhance slow oscillation activity.
However, the effect depends on the brain state, requiring the presence of
stable non-rapid eye movement sleep.
\end{abstract}
\normalsize
\section*{Introduction}
Sleep slow oscillations (SOs) of $<1$Hz hallmark slow-wave
sleep (SWS) as they are the largest oscillatory events
(amplitude $>75\mu$V) recorded in the electroencephalogram
(EEG). They emerge from highly synchronized cortical
neuronal networks undergoing alternations between phases
of membrane depolarization together with higher firing
activity (up states) and phases of hyperpolarized membrane
potentials and neural quiescence (down states;
Sanchez-Vives and McCormick, 2000; Steriade et al.,
1993). SOs have been likewise observed in diverse animal
species and humans (Achermann and Borbely, 1997; M\"olle
et al., 2002; Rattenborg et al., 2011; Steriade, 2006). A
large number of studies have demonstrated different
functions of the SO, the most important of which is their
role in the consolidation of long-term memory (Diekelmann
and Born, 2010; M\"olle and Born, 2011), and their role in
the homeostatic regulation of synaptic connectivity (Tononi
and Cirelli, 2006).

The established functional importance of SOs has
attracted growing interest regarding their manipulation in
terms of both enhancement and suppression (Landsness
et al., 2009; Marshall et al., 2006; Van der Werf et al., 2009).
The former requires a possibility to trigger SOs, which has
been successfully attempted using transcranial direct current
stimulation (tDCS; Marshall et al., 2006), transcranial magnetic
stimulation (Massimini et al., 2007) or intracranial
electrical stimulation (Vyazovskiy et al., 2009). Rhythmic
sensory stimulation, to the best of our knowledge, has not
been tested as a tool to induce SOs in humans, although
acoustic stimuli are well known to induce K-complexes, which
are considered a forerunner of the SO (Cash et al., 2009;
De Gennaro et al., 2000; Riedner et al., 2011). Based on this
evidence and because of the simplicity of the approach, here
we probed the capacity of rhythmic stimulation in the 0.8-Hz
SO frequency to induce SOs in the human brain. Of particular
interest was whether such regular stimulation had the
capability to entrain endogenous SO rhythms to an external
drive. Effects were tested while subjects were awake,
   \clearpage 
\noindent
transited into sleep and during stable non-rapid eye movement
(NonREM) sleep. Of additional interest was the question
whether rhythmic 0.8-Hz stimulation would accelerate
onset of sleep and SWS. A similar effect was achieved in a
previous work by instrumental conditioning of the sensorimotor
rhythm (Hoedlmoser et al., 2008). Basically, starting
the stimulation already during wakefulness before sleep also
enabled a comparison between the effects of the stimulation
between wakefulness and sleep, which should reveal clues
as to a possible brain state dependence of the stimulation
effects.\\[2ex]
\section*{Materials and Methods}
\subsubsection*{Subjects, experimental design and procedures}
Ten healthy subjects (seven females, three males; mean
age = 22.3 $\pm$ 1.0 years; range = 18–26 years) participated
in the experiments. All participants were non-smokers, and
were not using any medication at the time of the experiment.
Prior screening ensured no history of neurological or psychiatric
disease. Participants were not allowed to ingest alcohol
on the day before experimental nights, and were asked to
refrain from caffeine 8 h before the scheduled sleeping time.
Moreover, they were instructed to get up at 07:00 hours and
not to take a nap during these days. Prior to the experiments,
subjects were accustomed to sleeping under laboratory
conditions during an adaptation night, including EEG recordings
and wearing of headphones (but without any stimulation).
The experiment was approved by the ethics committee
of the University of Lu\"ubeck, and subjects gave written
informed consent prior to participation.

Each subject was studied according to a within-subject
design on three experimental conditions (0.8-Hz stimulation,
random stimulation, and sham), with the respective experimental
nights separated by at least 5 days. The order of
conditions was balanced across participants. In the 0.8-Hz
stimulation condition sound bursts were presented with a
constant interstimulus interval (ISI) of 1.25 s, corresponding
to a frequency of 0.8 Hz as an approximate to the SO
frequency. In the random stimulation condition, sounds
occurred randomly, with ISIs ranging from 0.125 to 5 s,
excluding intervals between 0.5 and 2 s in order not to
overlap with effects of the 0.8-Hz stimulation. Random ISIs
were generated such that the average ISI was also 1.25 s in
this condition, and that in both stimulation conditions the
same total number of sounds (n = 4416) was presented
(which implies a higher amount of short ISIs during the
random stimulation condition). Acoustic stimulation commenced
2 min before lights were turned off (at 23:00 hours).
During this 2-min interval subjects lied in bed with eyes open
fixating a point at the ceiling. After lights off, subjects were
allowed to sleep and acoustic stimulation continued for a
further 90 min. The sham control condition comprised periodic
acoustic presentation only within the 2 min prior to lights
off. The EEG was continuously recorded throughout the
whole night until 07:00 hours, when the participants were
awakened.
\subsubsection*{Acoustic stimulation}
The stimuli were bursts of pink $1/f$ noise of 50 ms duration,
with a 5-ms rising and falling time, respectively. Pink instead
of white noise was used because it sounds softer and is
therefore more comfortable to hear. Sound volume was
measured and calibrated prior to each experimental night
using a Voltcraft sound level meter SL-400 (Conrad Electronic
SE, Hirschau, Germany) to 60 dB SPL, measured
directly at the in-ear headphone. Stimuli were presented
binaurally via RP-HJE170 in-ear headphones (Philips,
Amsterdam, the Netherlands).
\subsubsection*{Sleep EEG recordings and polysomnography}
The EEG was recorded with a Neurofax EEG-9200 (Nihon
Kohden, Tokyo, Japan) from 19 channels (extended 10–20
system, Fp1, Fp2, F7, F3, Fz, F4, F8, T3, C3, Cz, C4, T4, T5,
P3, Pz, P4, T6, O1, O2) referenced to C3 and C4 using an
Easycap (Easycap GmbH, Herrsching, Germany) and Ag-
AgCl ring electrodes. Impedances were kept $<$5 k$\Omega$. Signals
were filtered between 0.08 and 120 Hz and offline
re-referenced to the averaged signal from mastoid electrodes
(M1, M2). Vertical and horizontal eye movements (VEOG,
HEOG) as well as electromyogram from the chin (EMG) were
obtained for standard polysomnography and for artefact
detection. All recordings were sampled at 500 Hz and stored
for later offline analyses.
Electroencephalogram (at C3 and C4), electrooculogram
(EOG) and EMG recordings were used for offline scoring of
sleep by two experienced raters who were blinded with
regard to the experimental condition. Scoring was done for
subsequent 30-s recording epochs according to standard
criteria (Rechtschaffen and Kales, 1968). Total sleep time
and time spent in the different sleep stages (wake; sleep
stages 1–4; SWS, i.e. sum of sleep stages 3 and 4; REM
sleep) were determined for the total nights as well as for the
90-min periods of acoustic stimulation (and for corresponding
periods of the sham condition). Also, sleep onset latency (first
occurrence of stage 1 sleep followed by stage 2 sleep, with
reference to lights off) and latency of SWS and REM sleep
(with reference to sleep onset) were determined. Prior to
scoring, the EEG and both EOG channels were low-pass
filtered at 30 Hz, and EMG channels were high-pass filtered
at 5 Hz. Stimuli evoking arousals or awakenings (as judged
by visual inspection) were marked in order to discard them
from averaging analyses.
\subsubsection*{EEG spectral analysis}
Analyses were performed with Spike2 software version 7
(Cambridge Electronic Design, Cambridge, UK) and Brain
Vision Analyser 2 (BrainProducts, Munich, Germany). All
   \clearpage\noindent 
EEG signals were pre-filtered between 0.15 and 30 Hz.
Beginning with the 2-min wake interval before lights off, a
Fast Fourier Transformation (Hanning window, 16 384 data
points) was calculated on a 33-s window that was moved in
10-s steps in time for a total of 32 min. Analysis was limited to
the 32-min time interval, as individual sleep courses became
highly divergent with ongoing sleep resulting in large interindividual
variance. To obtain the time course of the activity in
the SO (0.5–1 Hz), slow-wave activity (SWA; 0.5–4 Hz),
theta (4–8 Hz), slow (9–12 Hz) and fast spindle (12–15 Hz)
bands, for each 33-s window the mean spectral power within
the respective frequency band was calculated.
       \subsubsection*{Auditory-evoked potentials (AEPs)}
To assess AEP responses for the two stimulation conditions,
the EEG signals were averaged with reference to stimulus
onset. For the random stimulation condition, stimuli were
additionally divided into `overlapping' and `non-overlapping' 
stimuli, depending on whether or not the stimulus was
separated by more than 1.25 s from the previous and
following stimulus. As the non-overlapping category included
the lowest absolute number of stimuli, for respective comparisons
stimulus subsets of equal size were randomly drawn
from the category of overlapping stimuli as well as from the
set of stimuli in the 0.8-Hz stimulation condition. Signals were
averaged for a 1.2-s window including a 0.1-s pre-stimulus
onset baseline. To normalize responses, the average
potential during this baseline was set to zero. In the same
way, slow and fast spindle activities were averaged with
reference to stimulus onset. Prior to this analysis the EEG
signal was filtered in the respective frequency bands
(9–12 Hz, 12–15 Hz), down-sampled to 100 Hz, and the root
mean square signal was determined. The baseline normalization
was performed as described for AEPs.
   \subsubsection*{SO detection and statistical analyses}
Analyses of SOs were restricted to periods of SWS during the
90-min stimulation period and during the corresponding
period of the sham condition. Offline detection of SOs was
performed in all EEG channels according to a custom-made
algorithm described previously (M\"olle et al., 2002). In brief,
the EEG was band-pass filtered between 0.15 and 30 Hz,
and down-sampled to 100 Hz. For the identification of large
SOs a low-pass filter of 3.5 Hz was applied. Then, negative
and positive peak potentials were derived from all intervals
between consecutive positive-to-negative zero crossings (i.e.
one negative and one positive peak between two succeeding
positive-to-negative zero crossings). Only intervals with
durations of 0.8–2 s (corresponding to a frequency of 0.5–
1.25 Hz) were included. A SO was identified as such only if
both absolute negative and negative-to-positive peak potentials
were larger than 1.5 times the respective average.
Averages of original EEG potentials in a 3-s window $\pm$1.5 s
around the peak of the negative half-wave of all identified
SOs were calculated. To examine whether rhythmic acoustic
stimulation enhanced the occurrence of SOs in trains of
several succeeding oscillations, auto-event correlation analyses
were performed using 9-s intervals with 4.5-s offset and
a bin-size of 0.1 s. The histograms were referenced to the
negative half-wave peaks of the SO.

Statistical differences between experimental conditions
were assessed using analyses of variances (anova) and
paired t-tests. We concentrate here on results from t-tests
that are reported only after respective anova indicated
significance for the main or interaction effects of interest. A
P-level $<$0.05 was considered significant.
\section*{Results}
\subsubsection*{\mbox{Rhythmic 0.8-Hz stimulation delays sleep onset}}
Fig. 1 shows the average time subjects needed to transit
between stages, beginning from wakefulness to the first
occurrence of stage 1 sleep, from stage 1 to stage 2 sleep,
and from stage 2 sleep into SWS. Contrary to expectations,
in the 0.8-Hz stimulation condition subjects needed significantly
more time to reach stage 1 compared with both the
random stimulation (P $<$ 0.01) and the sham (P $<$ 0.05)
condition, indicating a delayed sleep onset. The transition
from stage 1 to stage 2 sleep was not affected by 0.8-Hz
stimulation, although it revealed to by delayed with random
stimulation (P $<$ 0.05, compared with sham). Differences in
the transition time from state 2 sleep into SWS were not
significant (P = 0.78 versus random stimulation; P = 0.37
versus sham).
\\
\includegraphics[width=0.49\textwidth]{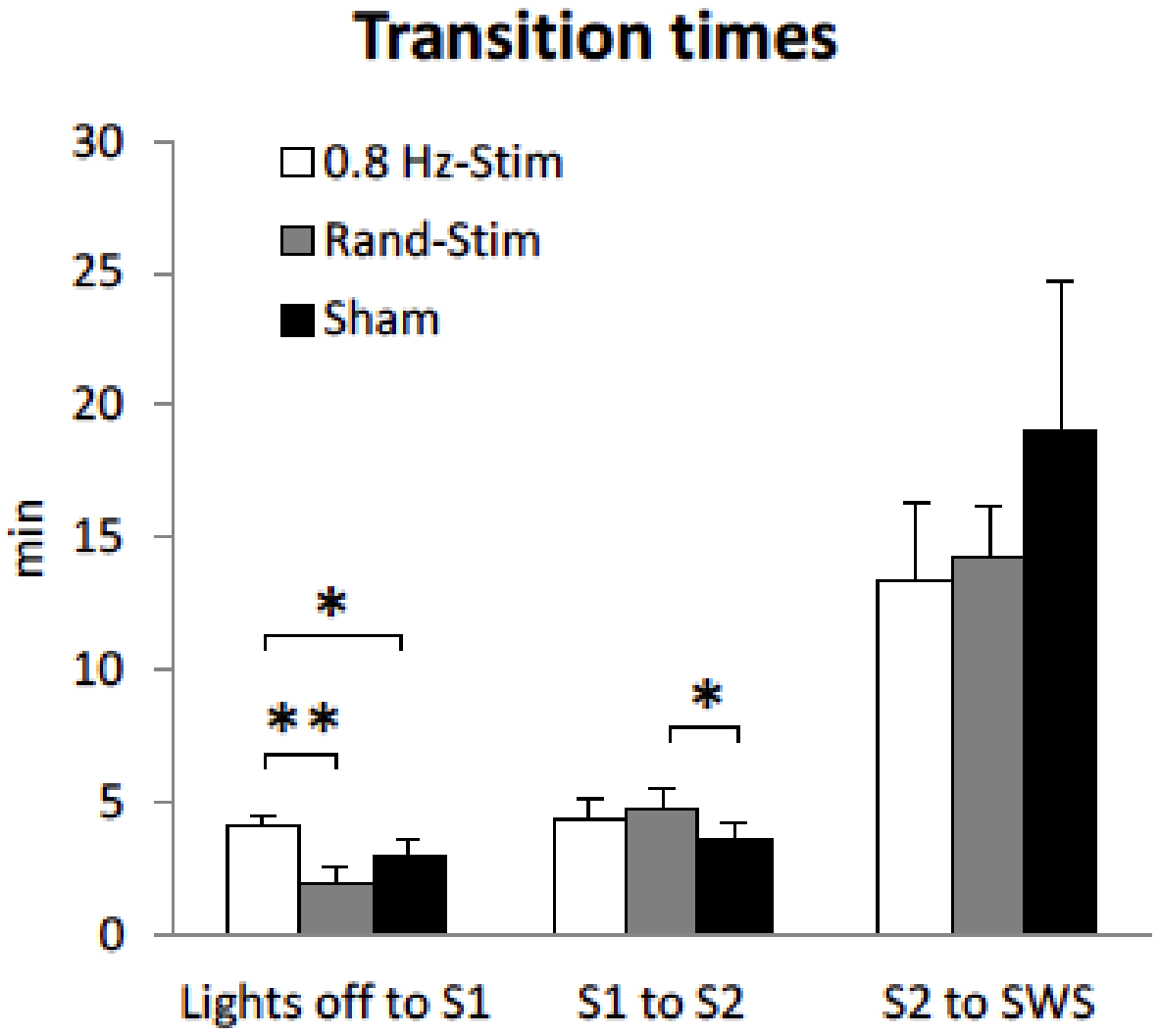}
\\
{\bf Figure 1.} 
{\small 
Mean ($\pm$SEM) transition times for the three stimulation
conditions (0.8-Hz stimulation, random stimulation, sham) to reach
sleep stage 1 (S1) from wakefulness (left), stage 2 sleep (S2) from
S1 (middle), and slow-wave sleep (SWS) from S2 (right). `Wakefulness' 
refers to the time when lights were turned off, and only transitions
to S1 were considered that were followed by S2. $^*P < 0.05$,
$**$P $<$ 0.01, for paired $t$-tests.}
        \clearpage\noindent 
\includegraphics[width=0.49\textwidth]{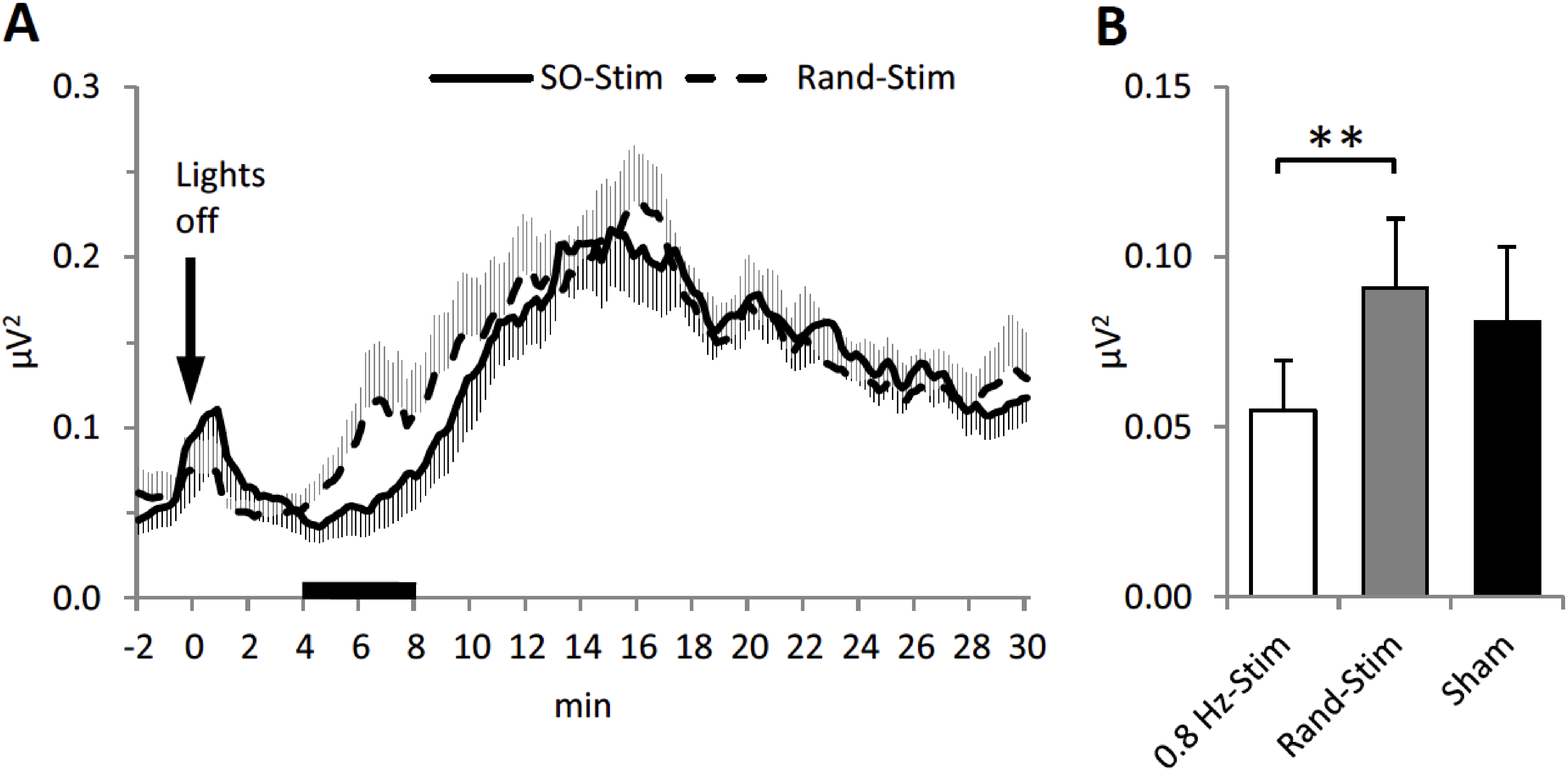}
\\
{\bf Figure 2.} {\small  Mean ($\pm$SEM) time course (a) of
spindle power (12–15 Hz, at Cz), with reference
to lights off (0 min) for the 0.8-Hz stimulation
condition (solid line) and for the random
stimulation condition (dashed line). The black
horizontal bar indicates an interval 4–8 min
after lights off where spindle power differed
significantly (P $<$ 0.01) between conditions. (b)
Mean ($\pm$SEM) spindle power at Cz for the 0.8-
Hz stimulation (white), random stimulation
(grey) and sham (black) stimulation conditions
4–8 min after lights off. **P $<$ 0.01 for paired
t-test.}
\\[6ex]
\includegraphics[width=0.49\textwidth]{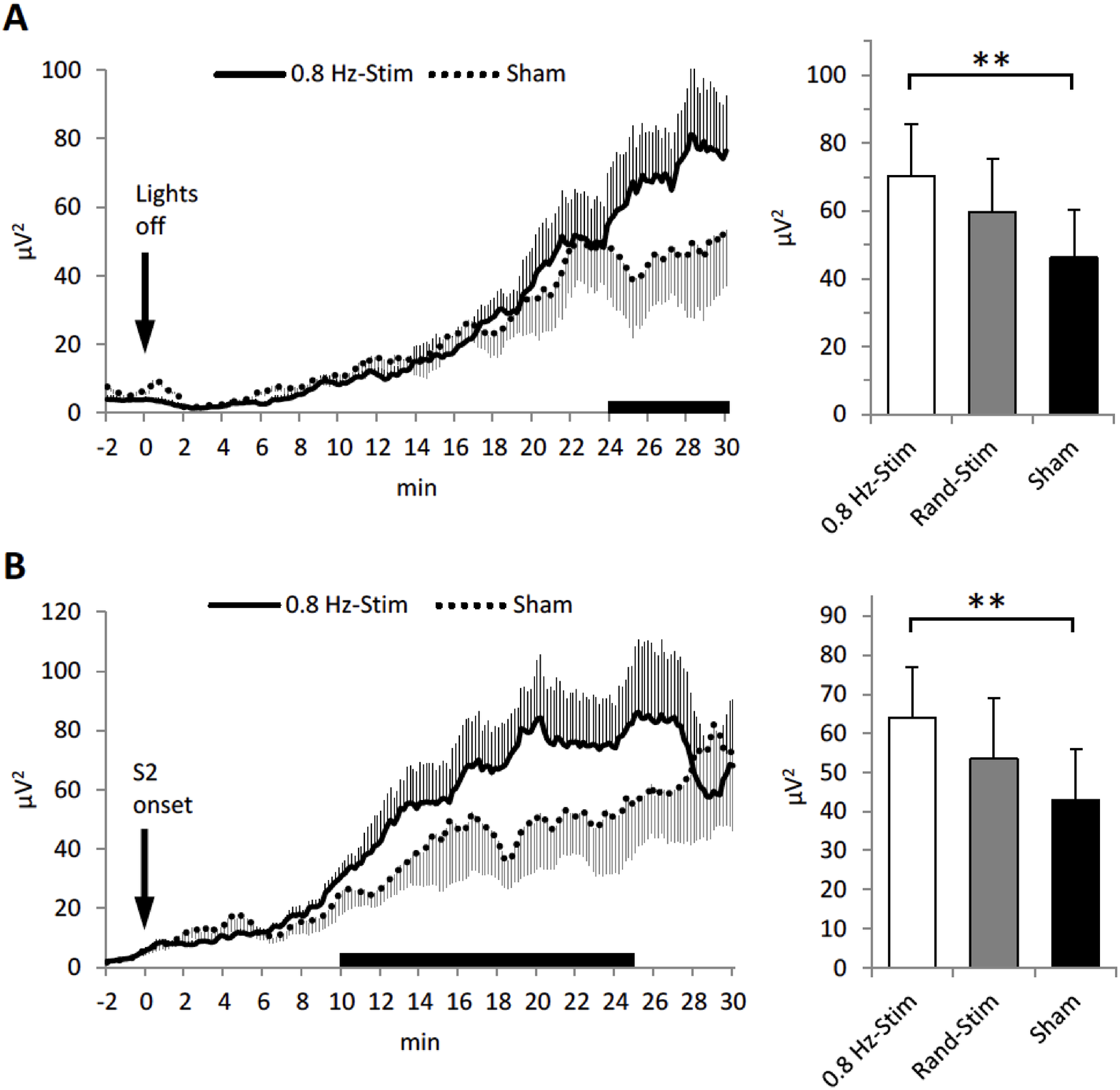}
\\
{\bf Figure 3.} {\small  Mean ($\pm$SEM) time course (left
panels) of SO power (0.5–1 Hz, at Fz) averaged
with reference to lights off (a) and with
reference to the onset of sleep stage 2 (b), for
the 0.8-Hz stimulation condition (solid line) and
for the sham condition (dotted line). Black
horizontal bars indicate intervals where SO
power differed significantly (P $<$ 0.01) between
conditions. Mean ($\pm$SEM) SO power for these
intervals is indicated for the 0.8-Hz stimulation
(white), random stimulation (grey) and sham
(black) stimulation conditions in the right panels.
**P $<$ 0.01 for paired t-test.
of 0.8-Hz stimulation on SWA were less consistent than
those on SO activity and overall revealed only marginal
significance. Effects on theta activity as well as on slow
spindle activity remained non-significant.}
\vspace{10ex}
\noindent
Analyses of fast spindle power (12–15 Hz) confirmed that
stimulation, and particularly 0.8-Hz stimulation, delayed the
occurrence of stable NonREM sleep (Fig. 2a). A few minutes
after subjects were allowed to sleep, fast spindle power
started to increase. However, this increase was delayed
during rhythmic 0.8-Hz stimulation. Thus, spindle power (at
Cz, averaged time-locked to lights off) was reduced during
0.8-Hz stimulation, most consistently if compared with
random stimulation (P $<$ 0.01; Fig. 2b).
        \subsubsection*{0.8-Hz stimulation enhances SO activity once stage S2 sleep has manifested}
Contrary to our expectation, 0.8-Hz stimulation, compared
with random stimulation and sham, did not affect power in the
SO frequency band (0.5–1 Hz) during the waking period
before lights off as well as thereafter in the beginning of the
sleep period (Fig. 3a). However, the rhythmic 0.8-Hz stimulation
had an impact once the subject advanced into
NonREM sleep stage 2. Averaging SO power time-locked
to the onset of sleep stage 2 revealed that 0.8-Hz stimulation
produced a distinct increase in SO power 10–25 min later,
with this effect coinciding with the occurrence of SWS
(Fig. 3b). The effect was most consistently observed when
compared with the sham condition (P = 0.011, for analyses
with reference to lights off, Fig. 3a; P = 0.004, for analyses
with reference to the onset of stage 2 sleep, Fig. 3b).
Changes in sleep onset latency during the 0.8-Hz stimulation
condition (with reference to sham stimulation) and SO power
were not correlated (r = 0.266, P = 0.458), excluding that
increases in SO power were an immediate consequence of
the delaying effect of the stimulation on sleep onset. Effects
        \clearpage\onecolumn\noindent
of 0.8-Hz stimulation on SWA were less consistent than
those on SO activity and overall revealed only marginal
significance. Effects on theta activity as well as on slow
spindle activity remained non-significant.
\subsubsection*{Auditory stimulation modulates slow and fast spindle
activity}
Averaged AEPs to the stimulation were determined separately
for the 0.8-Hz stimulation and the random stimulation
condition, with the responses for the random stimulation
condition additionally separated to stimuli that were or were
not separated by more than 1.25 s from the previous and
following sound (`non-overlapping' versus `overlapping' 
responses). Consistent with previous studies (e.g. Colrain
and Campbell, 2007), AEPs during stage 2 sleep revealed a
positive component about 200 ms post-stimulus onset followed
by a double-peaked negative component 300–600 ms
post-stimulus onset (Fig. 4, upper panel). The two peaks of
the latter component complex tended to merge into a single
broad hyperpolarization during SWS, which was then followed
by a depolarization at 900 ms post-stimulus. In fact,
during the SWS this late negative-to-positive AEP complex
(300–900 ms post-stimulus) showed some similarity with a
SO. Generally, the AEP potential components were smallest
for the `overlapping' responses, as compared with the `nonoverlapping' 
responses and with the responses to the 0.8-Hz
stimulation (see Fig. 4 for statistical comparisons), reflecting
the refractoriness of the AEP with shorter ISIs (Durrant and
Boston, 2006). AEPs in the random stimulation condition
\centerline{\includegraphics[width=0.75\textwidth]{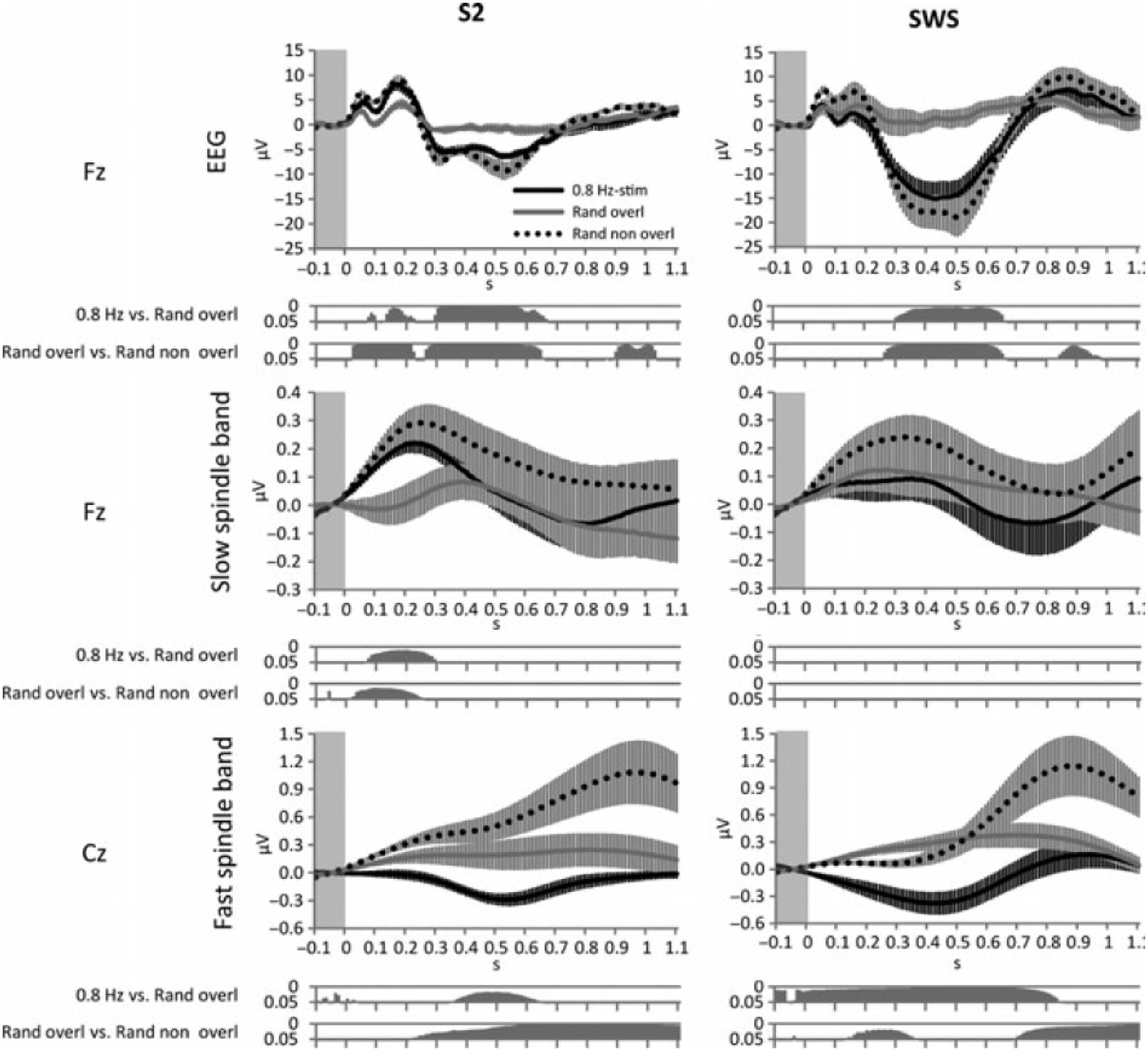}}
\\
{\bf Figure 4.} {\small AEPs (mean $\pm$ SEM) derived from 0.8-Hz stimulation (black line) and random stimulation conditions (grey line for overlapping and
dotted line for non-overlapping stimuli) and categorized by sleep stage S2 (left column) and slow-wave sleep (SWS; right column), as well as
conventional electroencephalogram (EEG) band (0.15–30 Hz, top row), slow spindle band (9–12 Hz, middle row) and fast spindle band (12–
15 Hz, bottom row). Bottom lines indicate point-wise statistical comparison between the 0.8-Hz stimulation condition and the overlapping
random stimulation (Rand overl), and between the overlapping and non-overlapping (Rand non-overl) stimuli of the random stimulation
condition. As the non-overlapping stimuli included the lowest number of stimuli, for respective comparisons stimulus subsets of equal size were
randomly drawn from the category of overlapping stimuli as well as from the set of stimuli in the 0.8-Hz stimulation condition. Vertical grey bars
indicate intervals used for baseline normalization. AEPs during S2: n = 211.8 $\pm$ 13.8; and during SWS: n = 194.9 $\pm$ 18.5.}
      \clearpage\twocolumn\noindent
were also significantly smaller than those during the 0.8-Hz
stimulation condition, when AEPs were averaged across all
stimuli (overlapping and non-overlapping stimuli of the
random stimulation condition, and all stimuli of the 0.8-Hz
stimulation condition; Fig. S1).
Root mean square slow spindle activity averaged with
reference to stimulus onset showed a maximum shortly
before the AEP negativity 300–600 ms post-stimulus, but did
not differ for the different stimulus types during SWS
(overlapping, non-overlapping random stimulation, 0.8-Hz
stimulation; Fig. 4). Fast spindle activity was suppressed
during the AEP negativity (300–600 ms post-stimulus), particularly
during the 0.8-Hz stimulation condition (see Figs 4
and S1 for statistical comparisons). This decrease in the
0.8-Hz stimulation condition was also significant when
compared with responses averaged across all stimuli of the
random stimulation condition (Fig. S1). Fast spindle activity
was enhanced during late AEP positivity (900 ms poststimulus),
in particular after non-overlapping stimuli of the
random stimulation condition. This late increase in fast
spindle activity was also significant when all stimuli of the
random stimulation condition were compared with the 0.8-Hz
stimulation condition (Fig. S1).
\subsubsection*{\normalsize \bf SOs are modulated and entrained by 0.8-Hz stimulation}
Fig. 5 depicts average SOs identified during epochs of SWS
occurring in the 90-min period of stimulation and during the
corresponding periods of the random stimulation and the
sham conditions. Notably, the 0.8-Hz stimulation did not
significantly change the number of detected SO events
during the 90-min stimulation period (n = 779.2 $\pm$ 124.6
versus random stimulation n = 730.9 $\pm$ 54.2, sham
n = 732.0 $\pm$ 123.5, P $>$ 0.661), underlining that the effect
of the stimulation was primarily on the temporal entrainment
of SOs. Averaging was performed with reference to the
negative half-wave peak of the SOs. Comparison of the SOs
in the three stimulation conditions shows a significantly
stronger depolarization of the positive depolarizing up states
before and after the negative half-wave of a SO event for
the 0.8-Hz stimulation condition, in comparison with both the
random and sham conditions during intervals of greatest
difference (see Fig. 5 for statistical comparisons). To analyse
the occurrence of trains of SOs in the different
stimulation conditions, we calculated auto-event correlation
histograms that visualize the timing between successive
SOs. These auto-event correlation histograms indicated that
rhythmic acoustic 0.8-Hz stimulation indeed induced more
regular trains of SOs during SWS (Fig. 6). This was
apparent by significant (P $<$ 0.05) increases in the frequency
of SO peaks around time points being multiples of 1.25 s
during the 0.8-Hz stimulation condition, i.e. the emerging
SOs adapted to the external drive. Again this entraining
effect of 0.8-Hz stimulation on the SOs was significant in
comparison with both the random stimulation and the sham
conditions.
\\
\centerline{\includegraphics[width=0.35\textwidth]{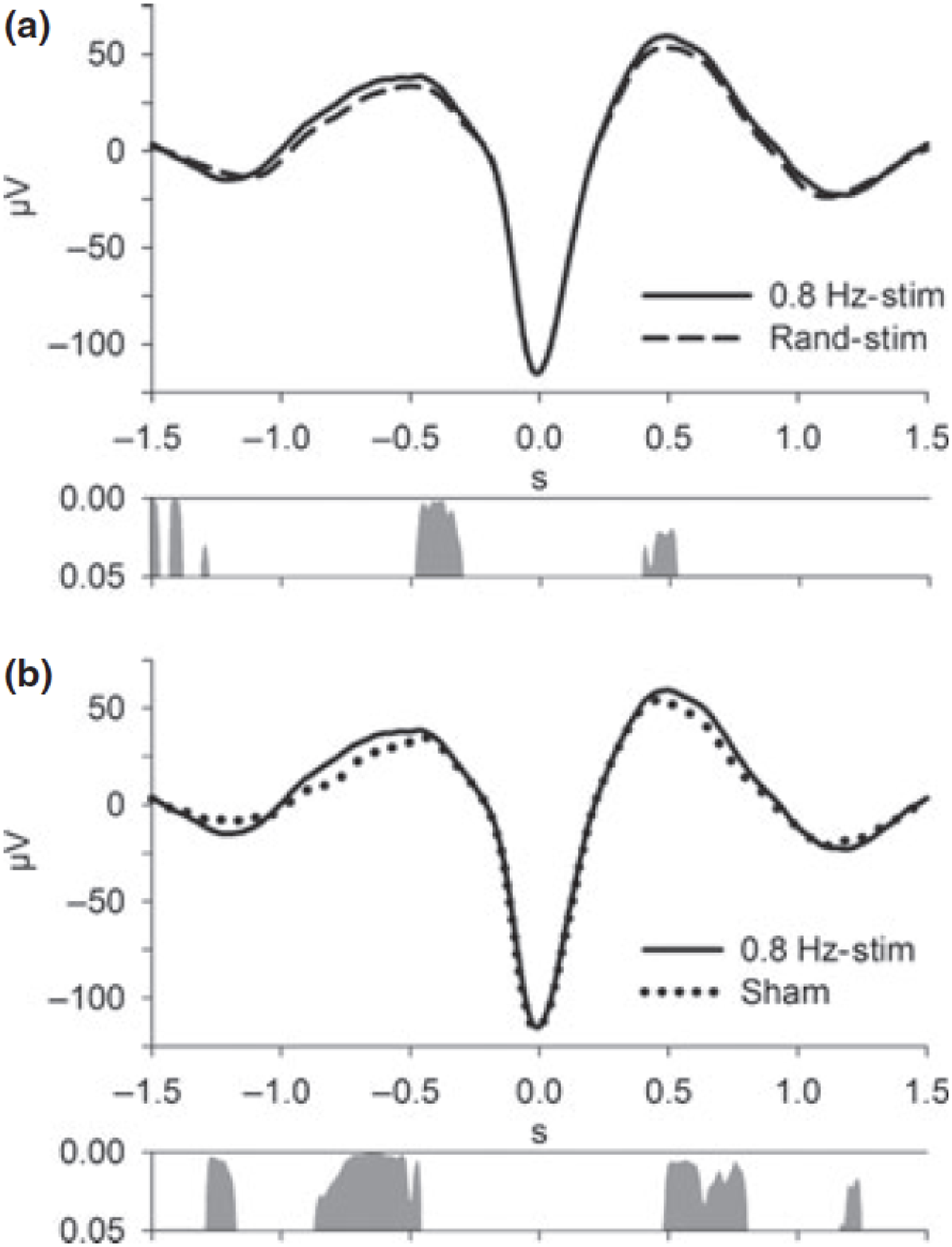}}
{\bf Figure 5.} {\small Mean SO (at Fz) during SWS periods of the 90-min
stimulation interval for (a) the 0.8-Hz stimulation condition (solid line)
and random stimulation (dashed line), and (b) 0.8-Hz stimulation and
sham conditions (dotted line). Differences in the potential level are
indicated by point-wise statistical comparisons positioned below each
average. SEMs were generally $<$3 lV and are not shown because at
the selected scaling they are not discernible.}
\subsubsection*{\bf Sleep architecture remained unchanged during and after
stimulation}
An analysis of the sleep stage distribution following the first
90 min was performed to examine whether continuing effects
were present and affected remaining sleep. Table 1 lists sleep
parameters for the 90-min period of stimulation as well as for
the remaining sleep epoch, and did not indicate any difference
between the stimulation conditions in time spent in the different
sleep stages for both periods, i.e. during and after stimulation
(P $>$ 0.096, for all comparisons). Also, number of arousals did
not significantly differ between conditions, excluding that
auditory stimulation disturbed sleep.
\section*{Discussion}
Our data indicate that rhythmic acoustic stimulation with a
slow 0.8-Hz frequency mimicking the frequency of natural
EEG SOs does not enhance SO activity in the waking brain,
but leads to a distinct delay in the onset of sleep, which was
paralleled by suppression of spindle power. However,
periodic 0.8-Hz stimulation increases spectral power in the
SO band during NonREM and SWS, although there was
no earlier onset of SWS. Averaging of AEPs and, in
parallel, evoked spindle activity revealed a stimulus-induced
\mbox{modulation of fast spindle activity reminiscent to that during}
          \clearpage\noindent
\onecolumn
\centerline{\includegraphics[width=0.499\textwidth]{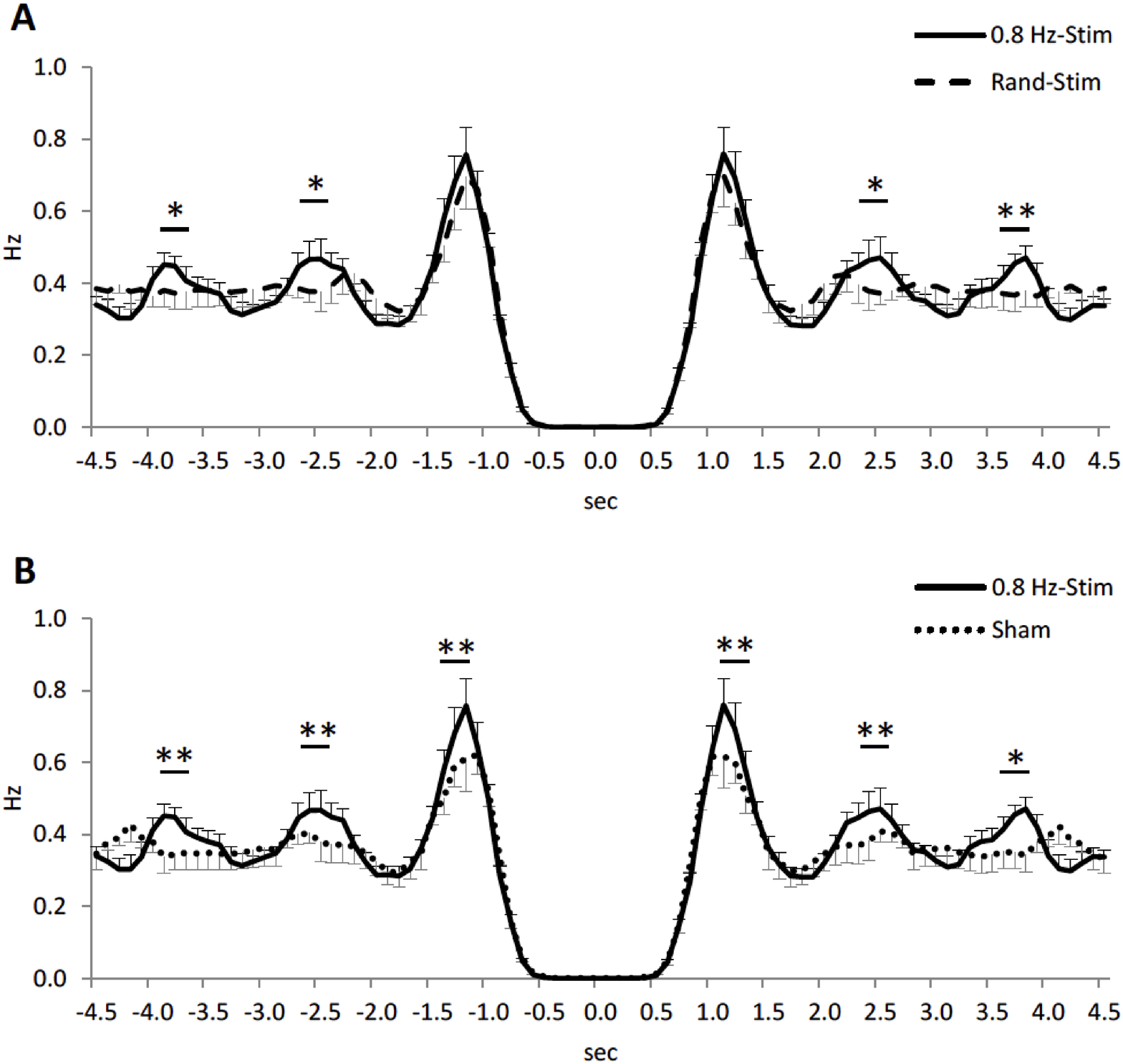}}
\noindent
{\bf Figure 6.} \footnotesize \small
Auto-event correlation of detected SOs (mean $\pm$ SEM) for recordings from Fz during SWS periods of the 90-min stimulation interval,
for the 0.8-Hz stimulation condition (solid lines), and (a) the random stimulation condition (dashed line) and (b) the sham condition (dotted line).
Negative peaks of detected SO events were used to perform auto-event correlation analyses (for 9-s windows around negative half-wave peaks,
0.1-s bin-size). The x-axis indicates time intervals between successive SO events; the y-axis indicates the rates of SO events occurring at a
given time interval. Thus, y-values represent the likelihood of a SO event occurring at a specific time before or after an identified SO event (as
defined by its negative peak at `0 s'). Black bars denote 0.3-s intervals centrally positioned at multiples of 1.25 s, corresponding to the 0.8-Hz
rhythm of acoustic stimulation. $*P < 0.05, **P < 0.01, $ for paired t-tests. Note the first peak in the auto-event correlation histogram (at 1.25 s) is
identical for both the 0.8-Hz stimulation and random stimulation conditions, reflecting that also randomly presented stimuli evoked two succeeding
SOs. However, the succeeding peaks of the histogram at 2.5 and 3.75 s were significantly higher for the 0.8-Hz stimulation condition
than the random stimulation condition, indicating that the 0.8-Hz stimulation induced longer trains of succeeding SOs.\\
\\[1ex]
\mbox{\bf Table 1 Sleep stage distribution for different time intervals and conditions}\\
\begin{tabular}{|c|c|c|c|c|c|c|}\hline
& \multicolumn{3}{|c|}{Stimulation period}& \multicolumn{3}{|c|}{Remaining sleep}\\\hline
Parameter & 0.8-Hz Stim &Rand stim &Sham &0.8-Hz Stim& Rand stim &Sham\\\hline
TST (min) &90.00 $\pm$ 0.0 &89.70 $\pm$ 0.1 &90.00 $\pm$ 0.0 &387.90 $\pm$ 4.0 &386.40 $\pm$ 4.9 &391.95 $\pm$ 1.2\\\hline
W (\%)& 4.56 $\pm$ 0.6 &3.45 $\pm$ 1.4 &3.56 $\pm$ 0.6 &1.28 $\pm$ 0.5 &0.93 $\pm$ 0.3 &0.97 $\pm$ 0.3\\\hline
S1 (\%) &5.61 $\pm$ 0.7 &5.63 $\pm$ 1.1& 6.28 $\pm$ 0.8 &2.92 $\pm$ 0.7 &2.22 $\pm$ 0.4 &1.97 $\pm$ 0.5\\\hline
S2 (\%) &46.56 $\pm$ 6.0 &43.65 $\pm$ 3.0 &50.22 $\pm$ 6.0 &55.94 $\pm$ 1.8 &52.61 $\pm$ 2.0 &58.08 $\pm$ 2.0\\\hline
SWS (\%) &39.06 $\pm$ 6.1 &38.01 $\pm$ 3.3 &36.83 $\pm$ 5.9 &10.04 $\pm$ 1.4 &13.41 $\pm$ 1.3 &10.56 $\pm$ 1.4\\\hline
REM (\%) &1.11 $\pm$ 1.1 &4.47 $\pm$ 1.9 &0.00 $\pm$ 0.0 &23.34 $\pm$ 1.5 &23.92 $\pm$ 1.3 &23.04 $\pm$ 1.1\\\hline
Arousals (\%) &3.06 $\pm$ 0.49 &4.63 $\pm$ 0.72 &3.06 $\pm$ 0.56 &5.87 $\pm$ 0.54&6.73 $\pm$ 0.51 &5.68 $\pm$ 0.68\\\hline
\multicolumn{7}{|c|}{
\begin{minipage}{0.97\textwidth} \small
Mean percentage ($\pm$SEM) of time spent in different sleep stages for the three stimulation conditions (0.8-Hz stimulation, random stimulation,
sham) and two time intervals: the 90-min stimulation period (measured from lights off) and the remaining stimulation-free sleep period (until
awakening at 07:00 hours). There were no significant differences between conditions.\\
REM, rapid eye movement; S1, sleep stage 1; S2, sleep stage 2; SWS, slow-wave sleep; TST, total sleep time; W, wake.
\end{minipage}
}
\\\hline
\end{tabular}
%
\\[2ex]
\normalsize
spontaneous SOs specifically in the 0.8-Hz stimulation
condition. Amplitude and auto-correlation analyses of SOs
revealed that the 0.8-Hz stimulation not only increased the
depolarizing up phase of SOs, but effectively entrained these
oscillations to the 0.8-Hz rhythm of stimulation.

The effect of acoustic stimulation on the sleeping brain has
been thoroughly investigated. However, the majority of these
studies in humans either focused on amplitudes and
latencies of specific components of the AEP response to
assess information processing during sleep (Bastuji et al.,
2002; Campbell and Colrain, 2002; Dang-Vu et al., 2011), or
aimed at a disturbance of sleep by acoustic stimulation to
suppress specific sleep stages like SWS and to investigate
its consequence on, for instance, learning performances
during subsequent wakefulness (Landsness et al., 2009; Van
der Werf et al., 2009). In anaesthetized guinea pigs, regular
          \clearpage\twocolumn\noindent
sound stimulation produced an entrainment of SO activity in
thalamic neurons (Gao et al., 2009). Against this background,
the present study is, to the best of our knowledge,
the first to examine whether rhythmic stimulation can be used
to entrain brain EEG oscillatory phenomena like the SO
during sleep in humans.

In agreement with other experimental studies, our findings
show that acoustic stimulation during SWS evokes a specific
electrophysiological response, consisting of a strong hyperpolarization
after about 500 ms followed by a depolarization,
which is maximal at about 900 ms (Amzica and Steriade,
1998; Plihal et al., 1996; Riedner et al., 2011). A strong
hyperpolarization followed by a depolarization is characteristic
for the SOs that are detected as such by our algorithm. It
is well established that this evoked response is associated
with strong cortical synchronized activity and interacts with
the thalamo-cortical system to generate the K-complex
(Bastien and Campbell, 1994; Contreras and Steriade,
1995). K-complexes bear striking similarities with the SO in
morphology and generating mechanisms, although differences
may also exist between these phenomena (Amzica,
2010; Cash et al., 2009; De Gennaro et al., 2000; Riedner
et al., 2011). Thus, in light of the fact that the increase in SO
power during the 0.8-Hz stimulation condition was not
statistically different from that of random stimulation, it could
be argued that these effects on SO power were mainly driven
by evoked K-complexes to the sounds rather than by an
entrainment to the rhythmic 0.8-Hz stimulation. To clarify this
issue, we analysed AEPs that indeed revealed that for stimuli
presented during random stimulation with
short (`overlapping' 
ISIs, the late negative-to-positive component complex
bearing great similarity with the SO was significantly smaller
than in the AEP to the stimuli presented at 0.8 Hz, likely
reflecting the refractoriness of the AEP response with short
ISIs (Durrant and Boston, 2006). Also, AEPs averaged
across all sounds of the random stimulation condition
revealed on average smaller component amplitudes than in
the 0.8-Hz stimulation condition, especially during sleep
stage 2. However, parallel analysis of evoked fast spindle
activity revealed that the observed increase in SO activity
during the 0.8-Hz stimulation condition cannot be entirely
reduced to K-complexes (evoked at this specific ISI).
K-complexes are typically associated with a transitory
increase in fast spindle activity (Contreras and Steriade,
1995). In fact, such an increase was observed in response to
the sounds (about 900 ms post-stimulus) during random
stimulation, with this increase significantly exceeding that
during 0.8-Hz stimulation. By contrast, in the 0.8-Hz stimulation
condition the suppression of fast spindle activity during
the preceding hyperpolarization of the AEP (300–600 ms
post-stimulus) predominated (Figs 4 and S1). Thus, whereas
isolated random stimuli caused a steady increase in fast
spindle activity over the entire 1.1-s post-stimulus interval,
the sounds of the 0.8-Hz stimulation condition produced a
phase-dependent modulation of fast spindle activity quite
similar to that observed during spontaneous SOs (M\"olle and
Born, 2011; M\"olle et al., 2002). Moreover, similar to the
temporal pattern during spontaneous SOs, also slow spindle
activity during the 0.8-Hz stimulation condition (during stage
2 sleep) was significantly increased at the transition of the
AEP into the negative phase ($\sim$300 ms post-stimulus; M\"olle
et al., 2011). This differential patterning of fast and slow
spindle activity, which is specifically observed during the
0.8-Hz stimulation condition and which closely mimics the
temporal relationships between slow and fast spindles during
spontaneous SOs, strongly argues for the view that factors
other than K-complexes significantly contribute to the
entrainment of SO activity observed during 0.8-Hz stimulation.
The effect on SO amplitude per se being only of
moderate size might reflect habituation concurrently developing
with the periodic signal presentation. In demonstrating
that rhythmic acoustic stimulation can induce and entrain
sleep SOs, our data suggest the use of this approach in the
study of functions known to be promoted by SOs, such as the
consolidation of memory and the post-sleep facilitation of
encoding of new memories (D Antonenko, S Diekelmann, C
Olsen, J Born and M M\"olle, 2012, submitted; Marshall et al.,
2006; Van der Werf et al., 2009). Such studies may reveal
the induction of trains of SOs to be more critical for memory
processes than mere changes in SO amplitude (M\"olle et al.,
2011).

A main finding of our study is that the efficacy of 0.8-Hz
stimulation is state dependent. SO power was enhanced by
the rhythmic stimulation only after stable NonREM sleep
stage 2 had become manifest. No similar effects were
obtained during waking before sleep, and the tone stimulation
also did not shorten sleep latency, which diverges from a
previous study (Bohlin, 1971), although that study used much
longer ISIs (varying between 20 and 40 s). A comparable
dependency of the effects of tone stimulation on the brain
state has been shown for the AEP showing characteristic
changes in its waveform (and frequency content) when the
brain transits from wakefulness to light sleep and SWS
(Campbell and Colrain, 2002; Cote, 2002). Here we revealed
a brain state dependence specifically in terms of predominant
EEG rhythm. The failure of 0.8-Hz stimulation to increase SO
activity in the waking brain, when the EEG is dominated by
faster frequencies, together with significant delay of sleep
onset resulting from the periodic 0.8-Hz stimulation implies
that the brain's susceptibility to an external drive is highly
sensitive to its current state of vigilance. Wakefulness either
does not allow an entrainment per se, or this specific brain
state is characterized by a resonance frequency disjoint to
the 0.8-Hz stimulation, which explains the delayed transition
into sleep stage 1 as the system's dynamics are perturbed.
The idea of a resonance effect induced by 0.8-Hz stimulation
is in particular coherent with the finding of higher accumulation
of SO power after SWS was reached. However,
although suggesting a brain state dependency of the stimulation
effect, our data do not entirely rule out that other
factors, like circadian rhythm, added to the effects of 0.8-Hz
stimulation on EEG activity occurring selectively during sleep.
          \clearpage\noindent
The state dependence of the effects of 0.8-Hz stimulation on
SOactivity reported here has been similarly observed in recent
studies using tDCS oscillating at a frequency of 0.75 Hz. The
oscillating tDCS induced wide-spread endogenous SO activity
when applied duringNonREMandSWS, whereas the increase
in SO activity was marginal and locally restricted to the
prefrontal cortex when the stimulation was applied to the
waking brain (Kirov et al., 2009; Marshall et al., 2006). However,
the waking brain responded with an increased theta
activity to 0.75-Hz tDCS. The convergence of these findings
tempts to speculate about the concept of a resonance
frequency that characterizes the oscillatory EEG response to
rhythmic stimulation for the different brain states. In this view,
SWS is essentially characterized by a 0.8-Hz resonance
frequency, whereas wakefulness as well as light sleep at the
transition to deep sleep represent brain states resonating at
frequencies different from the 0.8-Hz stimulation frequency
applied here. The modulation of 12–15-Hz spindle activity
caused by the random stimulation suggests for the transitory
period of light sleep a faster resonance frequency above the
SO range, as the random stimulation contained a high
proportion of shorter ISIs. Consistent with this view, in a
previous study, instrumental conditioning of sensorimotor
rhythm in the 12–15-Hz frequency band effectively decreased
sleep onset latency (Hoedlmoser et al., 2008). Yet, such
assumptions are in need of experimental validation.
In conclusion, our results indicate that rhythmic acoustic
stimulation can be used to induce sleep SO activity, which
makes it indeed a promising and simple approach for the
investigation of putative sleep functions linked to the SO
rhythm.
\small
\\[1ex] {\bf Achnowledgment}
This work was supported by the Deut\-sche Forschungsgemeinschaft
(SFB 654 and Graduate School 235).
\\ {\bf Conflict of Interest}
\small
None of the authors has any conflict of interest. We confirm
that we are in compliance with the Journal Sleep Research
policy.\\[-5ex]
\small
\bibliographystyle{enum}

\paragraph*{Supporting Information}
\mbox{} \\
\includegraphics[width=0.45\textwidth]{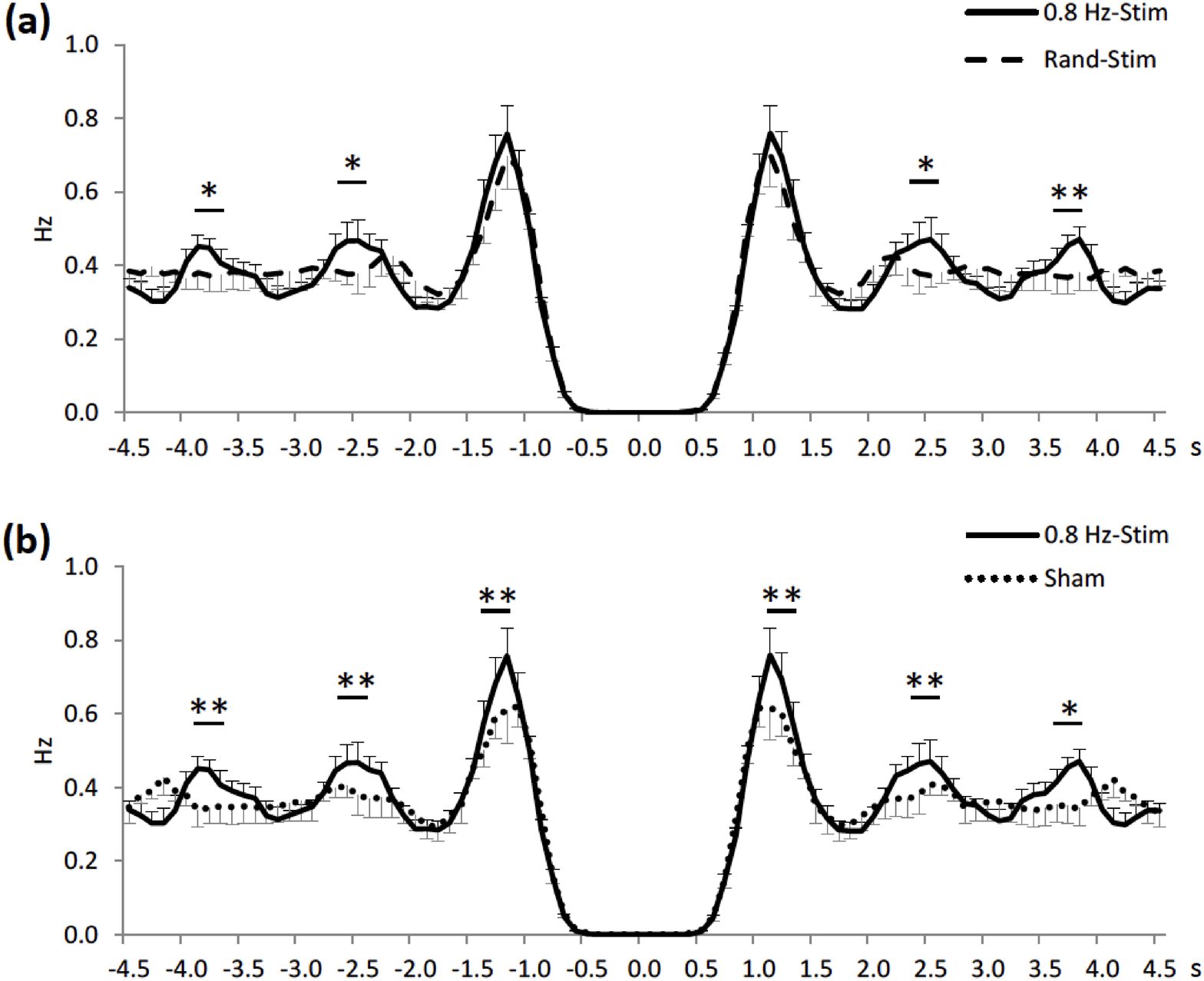}
\\
{\bf Figure S1.} {\small AEPs (mean $\pm$ SEM) derived from all auditory
stimuli during 0.8-Hz stimulation (black line) and random
stimulation conditions (grey line), and categorized by sleep
stage S2 (left column) and SWS (right column), as well as
conventional EEG band (0.15–30 Hz, top row), slow spindle
band (9–12 Hz, middle row) and fast spindle band (12–
15 Hz, bottom row). Additionally, point-wise statistical comparisons
are indicated below each AEP. Vertical grey bars
indicate intervals for baseline normalization. AEPs during S2
for 0.8-Hz stimulation: 2001.5 $\pm$ 254.5 and random stimulation:
1904.4 $\pm$ 130.0; and during SWS for 0.8-Hz stimulation:
1685.0 $\pm$ 261.4 and random stimulation: 1593.8 $\pm$ 138.7.}

\end{document}